\documentstyle[amssymb,aps]{revtex}

\begin{document}
\author{Qing-Yu Cai\thanks{%
qycai@wipm.ac.cn}}
\title{Information Erasure and Recover in Quantum Memory\thanks{%
Supported by the National Natural Science Foundation of China under Grant No
10304022.}}
\address{Wuhan Institute of physics and mathematics, The chinese Academy of Sciences,%
\\
Wuhan, 430071, China}
\maketitle

\begin{abstract}
We show that information in quantum memory can be erased and recovered
perfectly if it is necessary. That the final states of environment are
completely determined by the initial states of the system allows that an
easure operation can be realized by a swap operation between system and an
ancilla. Therefore, the erased information can be recoverd. When there is an
irreversible process, e.g. an irreversible operation or a decoherence
process, in the erasure process, the information would be erased
perpetually. We present that quantum erasure will also give heat dissipation
in environment. And a classical limit of quantum erasure is given which
coincides with Landauer's erasure principle.

\begin{description}
\item  PACS: 03.65.Bz, 03.67.Hk
\end{description}
\end{abstract}

The formalism of quantum mechanics allows the vast information content of a
quantum state to be efficiently processed, at a rate that cannot be matched
in real time by any classical means. Recently, a great deal of attention has
been focused on quantum computation.$^{[1-3]}$ Quantum computation is a
field of fundamental interest because we believe quantum information
processing machines can actually realized in nature. The elementary units of
the theory are quantum qubits. When memory in a quantum computer is scarce
(since any real physical resource is finite), erasure may plays an important
role in its own way since the quantum no-deleting principle states that the
linearity of quantum theory does not allow us to delete a copy of an
arbitrary quantum state perfectly.$^{[4-7]}$ As is well known, information
of an unknown classical bit can be erased perfectly.$^{[8-9]}$ In this
paper, we will show that information in quantum memory can be erased and
recovered perfectly if it is necessary. And the information will be erased
perpetually under an irreversible operation. In this case, information can
not be recovered. And the heat dissipation is inevitable.

In quantum theory, erasure an quantum state can be described, for an
arbitrary state $|\varphi >$, $^{[10]}$ 
\begin{equation}
\varepsilon |\varphi >\rightarrow |0>,
\end{equation}
where $\varepsilon $ is an erasure operation and $|0>$ is the standard
state. If the state is known, then one can prepare it to a standard state by
using a unitary operation. Consider a big number of qubits to be erased of a
quantum memory in quantum computer. These qubits should be treated as in the
unknown states. Suppose $\varepsilon $ is unitary and the state is unknown.
For two arbitrary states $|\varphi _{0}>$ and $|\varphi _{1}>$, we have 
\begin{eqnarray}
\varepsilon |\varphi _{0} &>&\rightarrow |0>, \\
\varepsilon |\varphi _{1} &>&\rightarrow |0>.
\end{eqnarray}
Since unitary transformations preserve the inner products, it must be that 
\begin{equation}
<\varphi _{0}|\varphi _{1}>=<0|0>.
\end{equation}
Clearly, $\varepsilon $ is nonunitary. The quantum mechanics postulate
states that the evolution of a closed quantum system is described by a
unitary transformation. Since $\varepsilon $ is nonunitary, then an erasure
process requires the system open. Without loss of generality, we assume that
the erasure process is to unitarily interact the state ($|\varphi _{0}>$ or $%
|\varphi _{1}>$) with an ancilla $|e>$,$^{[11]}$ in the two case one
obtaining 
\begin{eqnarray}
U|\varphi _{0} &>&|e>\rightarrow |0>|e_{0}>, \\
U|\varphi _{1} &>&|e>\rightarrow |0>|e_{1}>.
\end{eqnarray}
Taking inner product gives that 
\begin{equation}
<\varphi _{0}|\varphi _{1}>=<e_{0}|e_{1}>.
\end{equation}
Because $|\varphi _{0}>$ and $|\varphi _{1}>$ are arbitrary states, then the
final states of environment are completely determined by the initial states
of system. Since the erasure process can be realized by using a unitary
operation $U$ together with ancilla, it seems that an erasure process is
reversible. That is the case. But if one performs an irreversible operation,
e.g. measurement operation, on the ancilla, the erasure process becomes
irreversible immediately.

Since the final states of system is completely determined by the initial
states of system, then the operation $U$ can be consider as a swap
transformation $U_{S}$ if ancilla is in the standard state $|0>$, 
\begin{equation}
U_{S}|\varphi >|0>\rightarrow |0>|\varphi >.
\end{equation}
When one needs to use the erased information, he can using a swap operation
to recover it, 
\begin{equation}
U_{S}|0>|\varphi >\rightarrow |\varphi >|0>.
\end{equation}
However, in a physical erasure process, real environment is a heat
reservoir. And the final states of ancilla will be destroyed under the
decoherence, which leads this erasure process to irreversibility. That is, a
perpetual erasure operation can be realized by a swap transformation
together with an irreversible process that may be naturally completed by the
decoherence of environment. Consider the case of a single qubit in an
unknown state $|\varphi >=\alpha |0>+\beta |1>$, where $\alpha $, $\beta $
are arbitrary complex numbers and $|\alpha |^{2}+|\beta |^{2}=1$. Suppose
that ancilla is in the standard state $|0>$. Equation (8) tells us that a
swap operation 
\begin{equation}
U_{S}=\left( 
\begin{array}{llll}
1 & 0 & 0 & 0 \\ 
0 & 0 & 1 & 0 \\ 
0 & 1 & 0 & 0 \\ 
0 & 0 & 0 & 1
\end{array}
\right)
\end{equation}
can help the erasure process to be completed. Equation (8) can be rewritten
as 
\begin{equation}
\left( 
\begin{array}{llll}
1 & 0 & 0 & 0 \\ 
0 & 0 & 1 & 0 \\ 
0 & 1 & 0 & 0 \\ 
0 & 0 & 0 & 1
\end{array}
\right) \left( 
{\alpha  \choose \beta }%
\otimes 
{1 \choose 0}%
\right) =%
{1 \choose 0}%
\otimes 
{\alpha  \choose \beta }%
.
\end{equation}
Then the qubit is in a standard state after the swap transformation. The
ancilla is in the unknown state $|\varphi >=\alpha |0>+\beta |1>$. Clearly
,this is an reversible process. However, when we perform a measurement on
the ancilla in the basis $\{|0>,|1>\}$, the state $|\varphi >=\alpha
|0>+\beta |1>$ immediately collapses to a known state $|0>$ with a
probability $|\alpha |^{2}$ or $|1>$ with a probability $|\beta |^{2}$. The
information of the state $|\varphi >=\alpha |0>+\beta |1>$ can not be
recovered. A perpetual erasure process is realized.

In fact, an erasure operation can be realized by an irreversible operation
followed with a unitary operation. One can perform a measurement operation
to prepare the memory in a known state. Then the quantum memory can be
prepared in the standard state $|0>$ with a unitary operation. On this
occasion, information can not be recovered once it has been erased. Since
the quantum memory is open in the erasure process, there is energy exchange
between system and environment. In microworld, the minimal energy exchange
to perform a logical operation is of order $\hslash \omega $.$^{[12]}$ We
can reasonably assume that energy exchange between system and environment
when erasure one qubit is at least of order $\hslash \omega $. Suppose one
qubit is a two-level system which has one degree of freedom. Using theorem
of equipartition of energy, the average energy exchange when erasure one
qubit can be calculated 
\begin{eqnarray}
\stackrel{-}{\varepsilon } &=&\frac{\sum_{n=0}^{\infty }n\varepsilon
_{0}e^{-n\varepsilon _{0}/k_{B}T}}{\sum_{n=0}^{\infty }e^{-n\varepsilon
_{0}/k_{B}T}} \\
&=&\frac{\varepsilon _{0}}{e^{\varepsilon _{0}/k_{B}T}-1},
\end{eqnarray}
where $\varepsilon _{0}=\hslash \omega $. That is to say, when erasure one
qubit in the quantum memory, on average, there is $\stackrel{-}{\varepsilon }
$ heat dissipation in environment.

Consider the classical limit of the information erasure and recover in
quantum memory. Bennett has shown that all the logical irreversible
operations can be reconstructed by logical reversible operations.$^{[13]}$
Then the erased information can be recovered in classical computer. From
equation (13), when we accept that $\hslash \rightarrow 0$ in classical
limit, perpetually erasure one bit in classical computer will give heat
dissipation of order 
\begin{equation}
\stackrel{-}{\varepsilon }=k_{B}T,
\end{equation}
which coincides with Landauer's erasure principle.

In summary, we show that information erasure can be realized by using a
unitary operation together with an ancilla. The final states of ancilla is
completely determined by the initial states of the quantum memory. The
information can be recovered if it is necessary. When an irreversible
operation was performed, the erased information can not be recovered
forever. And this perpetual erasure operation is corresponding to a heat
dissipation in environment.

\section{references}

[1] Deutsch D and Jozsa R 1992, Proc. R. Soc. Lond. A $439$, 553

[2] Shor P W 1994, Proceedings of the 35th annual symposium on the
foundation of the computer science, IEEE Computer Society Press, Los
Alamitos, CA, p124

[3] Grover L K 1997, Phys. Rev. Lett. $79$ 325

[4] Pati A. K. and Braunstein 2000 Nature $404$ 164

[5] Feng S, Zheng S and Yim M 2002 Phys. Rev. A 65 042324

[6] Qiu D 2002 Phys. Rev A $65$ 052303

[7] Feng J, Gao Y F, Wang J S and Zhan M S 2002 Phys. Rev A $65$ 052311

[8] Landauer R 1961 IBM J Res. Dev. $5$ 183

[9] Bennett C. H. 1982 Int. J. Theor. Phys. $21$ 905

[10] Piechocinsha B 2000 Phys. Rev. A $61$ 062314

[11] Nielsen M A and Chuang I L 2000 Quantum Computation and Quantum
Information (Cambridge: Cambridge University Press)

[12] Gea-Banacloche J 2002 Phys. Rev Lett. $89$ 217901

[13] Bennett C H 1973 IBM J. Res. Dev. $17$ 525

\end{document}